\documentstyle[prd,preprint,aps]{revtex}
\begin{document}
\draft
\preprint{GTP-97-03}
\title{ The incident wave in Aharonov-Bohm scattering wavefunction }
\author
{$^{a}$Sahng-kyoon.Yoo and $^{b}$D.K.Park}
\address
{$^{a}$Department of Physics, Seonam University, Namwon 590-711, Korea \\
 $^{b}$Department of Physics, Kyungnam University, Masan 631-701, Korea}
\date{\today}
\maketitle
\begin{abstract}
It is shown that only the infinite angular momentum quantum states contribute
to the incident wave in Aharonov-Bohm (AB) scattering. This result is clearly 
shown by recalculating the AB calculation with arbitrary decomposition of 
summation over the angular momentum quantum numbers in wave function. 
It is motivated from the fact that the pole contribution in the integral 
representation used by Jackiw is given by only the infinite angular momentum 
states, in which the closed contour integration involving this pole gives just 
the incident wave. 
\end{abstract}

\indent Since Aharonov-Bohm(AB)\cite{aharonov59} discovered that particles in 
the field free region exhibit the quantum mechanical interference, known as the
AB effect, many physicists
\cite{ruijsenaars83,sakoda96,hagen90,jackiw90,berry80,takabayashi85,ouvry94,alvarez95,stelitano95} 
have studied in detail the 
scattering mechanism until now. There are, however, some disagreements in
the scattering wave function, in particular, in incident wave.

\indent Some authors\cite{ruijsenaars83,sakoda96} advocate the plane wave
as an incident wave, while others
\cite{aharonov59,jackiw90,berry80,takabayashi85,ouvry94,alvarez95}
derived the plane wave modulated by flux. This debate is still being in 
progress. In the former case, the commutativity in the two performances,
i.e., summation over the angular momentum states and the asymtotic limit
of Bessel function, in exact wavefunction must be tested, as pointed out
by Hagen\cite{hagen90}. Meanwhile, for the latter case, the plane wave
modulated by flux as an incident wave is derived by various method.
The incident wave and the scattering amplitude were calculated by
AB\cite{aharonov59} for the first time, by decomposing the wavefunction into
positive, negative and zero angular momentum states. (They have restricted
the ratio of flux to flux quantum from -1 to 1). They deduced the incident
wave exactly from the wavefunction, while the scattered wave for the 
non-forward direction in the asymptotic limit. Jackiw\cite{jackiw90} found
another method for the derivation of incident and scattered waves by using the
integral representation of Bessel function. He obtained the incident wave
from the contour enclosing the pole, and the scattered wave from the 
remaining straight line contour. The two methods mentioned above give
the exactly same results.

\indent In this paper, focusing on the latter case (modulated plane wave as an
incident wave), we investigate how the summation over angular momentum states
are related to the closed contour yielding the incident wave. We find that
only the infinite angular momentum states correspond to the pole in 
wavefunction, while all the other finite states represent the scattered
wave. This fact leads us to recalculate the AB wavefucntion by dividing the
summation not into positive, negative and zero component, as was done by AB,
but with the introduction of general decomposition, as will be described
below. We find that the deduced incident and scattered waves are same
regardless of the way of decomposition. This means that in AB scattering
the information on the incident wave cannot be involved in the finite
angular momemtum quantum states. It can be interpreted classically that
only the waves infinitely far away from the flux tube at origin do not scatter
at all, while the remaining ones are deflectd due to the AB potential. 
However, it is uncertain whether such a property is general one in all kinds 
of the 
long-ranged potentials such as Coulomb potential as well as AB potential. 
It is being in progress.

\indent Let us begin with the Schr\"{o}dinger equation representing a system
affected by AB potential in 2D polar coordinates:
\begin{equation}
\left[ \frac{\partial^2}{\partial r^2} + \frac{1}{r} 
\frac{\partial}{\partial r} + \frac{1}{r^2} \left( 
\frac{\partial}{\partial \phi} + i \alpha \right)^2 + k^2 \right] \psi
= 0,
\label{schrodinger}
\end{equation}
where $\alpha = -e \Phi / 2 \pi$, $e$ is the particle charge, $\Phi$ is the
magnetic flux, and $k$ is the wave number. We choose the gauge in which
$A_r = 0$ and $A_{\phi} = \Phi / 2 \pi r$.

\indent The regular solution of this equation is given by
\begin{equation}
\psi = \sum_{m= -\infty}^{m= \infty} (-i)^{ |m+ \alpha|} J_{|m+ \alpha|} (r') 
 e^{i m \phi},
\end{equation}
where $J_{\nu} (x)$ is the ordinary Bessel function of  $\nu$-th order and 
$r' = kr$. If one takes the flux
\begin{equation}
\alpha = N + \beta,
\end{equation}
where $N$ is an integer and $0 < \beta < 1$, then the summation can be 
decomposed into postive and negative parts, respectively:
\begin{equation}
\psi = \sum_{m= -N}^{\infty} (-i)^{ m+ \alpha} J_{m+ \alpha} (r') 
 e^{i m \phi}
 + \sum_{m=-N-1}^{- \infty} (-i)^{ - (m+ \alpha) } J_{- (m+ \alpha) } (r')
 e^{i m \phi}.
\label{abequation}
\end{equation}
The above summations over the angular momentum states can be performed by 
using the integral representation for the Bessel function\cite{gradshteyn80}
whose contour is given in figure 1:
\begin{equation}
J_{\nu} (x) = e^{ i \nu \pi /2} \int_{Cs} \frac{dz}{2 \pi}
e^{ -i x \cos z + i \nu z}.
\label{contour1}
\end{equation}
Inserting Eq.(\ref{contour1}) into Eq.(\ref{abequation}) and changing the 
variable $z \rightarrow -z$ in the second term, we get
\begin{eqnarray}
\psi & = & \left( \int_{C_{s}} - \int_{C_{-s}} \right) \frac{dz}{2 \pi} 
 e^{- ir' \cos z - i \alpha z} \sum_{m=-N}^{ \infty} e^{ i m (\phi -z)} 
\label{contour2}
\\
 & = & \int_{C} \frac{dz}{2 \pi} e^{i r' \cos (\pi -z)} 
 e^{i \alpha (\phi - \pi)}
 \frac{e^{-i \beta z}} {1 - e^{- iz}},
\label{contour3}
\end{eqnarray}
where the contour $C_{-s}$ is the mirror of $C_s$, and the contour $C$ is 
given in figure 2. Note that both the upper and lower contour in complex 
$z$-plane are shifted upward and downward by small positive quantity 
$\epsilon$. This is to avoid the poles in real axis raised in the course of
summation of Eq.(\ref{contour2}).
The contour $C$ can be divided into two parts: closed contour involving poles
in real axis ($C_1$) and straight line ($C_2$), which is drawn in figure 3. 
Jackiw obtained the incident wave from $C_1$ integration, 
while the scattered amplitude from $C_2$ integration.

\begin{picture}(230,270)(50,250)
\put (150,330) {\line(1,0){270}}
\put (290,330) {\line(0,1){150}}
\put (180,470) {\vector(0,-1){70}}
\put (180,400) {\line(0,-1){70}}
\put (180,330) {\vector(1,0){70}}
\put (330,330) {\vector(0,1){70}}
\put (330,400) {\line(0,1){70}}
\put (285,485) {Im $z$}
\put (430,330) {Re $z$}
\put (167,315) {$- \frac{3 \pi}{2}$}
\put (327,315) {$ \frac{\pi}{2}$}
\put (250,315) {\large{$C_s$}}
\put (180,270) {figure 1. Contour representing $J_{\nu} (x)$.}
\end{picture}

\indent Here, let us take only finite summation in Eq.(\ref{contour2}) to
explore the relation between the incident wave and the partial wave. 
In case of the first term, assuming that $M$ is large but finite, we have
\begin{equation}
\sum_{-N-1}^{M} e^{i m (\phi -z)} = e^{-i (N+1) (\phi -z)}
\frac{ \left[ 1- e^{ i (M+N+2) (\phi -z)} \right] }
{1- e^{i (\phi -z) }}.
\end{equation}
This is finite at $z = \phi$ as long as $M$ is finite.
This result is also true for the second term. Therefore,
if we exclude the infinite contribution from the summation, then the poles
in integrals do not appear. After all, only the infinite angular momentum
states do contribute to the incident wave in AB scattering. It leads us to
reconsider the AB calculation\cite{aharonov59}.

\begin{picture}(230,300)(50,190)
\put (130,350) {\line(1,0){270}}
\put (270,240) {\line(0,1){220}}
\put (200,360) {\vector(0,1){55}}
\put (200,245) {\vector(0,1){55}}
\put (200,415) {\line(0,1){40}}
\put (200,300) {\line(0,1){40}}
\put (200,360) {\line(1,0){110}}
\put (200,340) {\vector(1,0){110}}
\put (370,360) {\vector(-1,0){60}}
\put (310,340) {\line(1,0){60}}
\put (370,455) {\vector(0,-1){55}}
\put (370,340) {\vector(0,-1){55}}
\put (370,285) {\line(0,-1){40}}
\put (370,360) {\line(0,1){40}}
\put (410,350) {Re $z$}
\put (265,470) {Im $z$}
\put (275,365) {$\epsilon$}
\put (275,330) {$- \epsilon$}
\put (310,375) {\large{$C$}}
\put (200,353) {\line(0,-1){6}}
\put (370,353) {\line(0,-1){6}}
\put (170,330) {$\phi - \pi$}
\put (375,330) {$\phi + \pi$}
\put (180,210) {figure 2. Contour $C$ representing the Eq.(\ref{contour3}).}
\end{picture}

\begin{picture}(230,350)(50,-40)
\put (130,110) {\line(1,0){270}}
\put (270,0) {\line(0,1){220}}
\put (200,5) {\vector(0,1){175}}
\put (200,180) {\line(0,1){35}}
\put (205,120) {\line(1,0){105}}
\put (205,100) {\vector(1,0){105}}
\put (365,120) {\vector(-1,0){55}}
\put (310,100) {\line(1,0){55}}
\put (370,215) {\vector(0,-1){175}}
\put (370,40) {\line(0,-1){35}}
\put (370,120) {\line(0,1){60}}
\put (205,120) {\vector(0,-1){10}}
\put (205,110) {\line(0,-1){10}}
\put (365,100) {\vector(0,1){10}}
\put (365,110) {\line(0,1){10}}
\put (410,110) {Re $z$}
\put (265,230) {Im $z$}
\put (320,130) {\large{$C_1$}}
\put (180,170) {\large{$C_2$}}
\put (375,170) {\large{$C_2$}}
\put (275,125) {$\epsilon$}
\put (275,85) {$- \epsilon$}
\put (170,100) {$\phi - \pi$}
\put (375,100) {$\phi + \pi$}
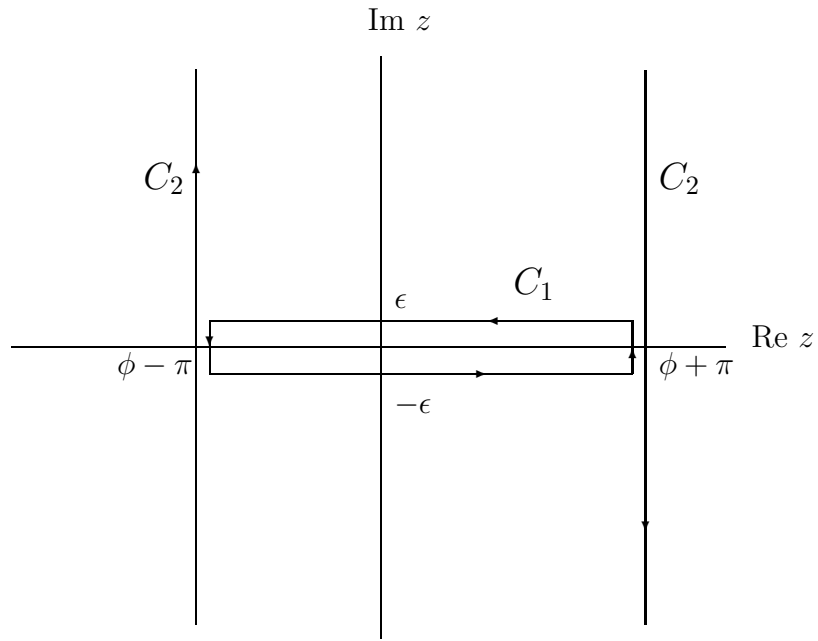
\put (180,-30) {figure 3. Equivalent contour with figure 2.}
\end{picture}

\newpage
\indent Now, we generalize the AB calculation as follows:

\indent The wavefunction of AB scattering is again given by 
Eq.(\ref{schrodinger}).
Unlike the AB calculation, we split $\psi$ into three parts:
\begin{eqnarray}
\psi_1 & = & \sum_{m=-N+M}^{\infty} 
(-i)^{ (m+ \alpha)} J_{(m+ \alpha)} (r') e^{i m \phi},
\\
\psi_2 & = & \sum_{m= -N-1-M}^{- \infty}
(-i)^{ - (m+ \alpha) } J_{- (m+ \alpha) } (r') e^{ i m \phi}
\nonumber \\
& = & \sum_{m=N+1+M}^{\infty} 
(-i)^{ (m- \alpha) } J_{(m- \alpha) } (r') e^{- i m \phi},
\\
\psi_3 & = & \sum_{-N-M}^{-N+M-1} 
(-i)^{ |m+ \alpha|} J_{|m+ \alpha|} (r') e^{i m \phi},
\end{eqnarray}
where $M$ is a large positive integer. We then follow the procedure of AB.
Note that $\psi_2$ is obtained from $\psi_1$
by replacing $\alpha$ by $- \alpha$ and $\phi$ by $- \phi$, and $\psi_3$ 
contains $2M$ terms unlike in AB. Let us consider first the cases of $\psi_1$ 
and $\psi_2$. The differential equation satisfied by $\psi_1$ is
\begin{equation}
\frac{d \psi_1}{dr'} = -i \cos \phi \psi_1 + \frac{1}{2} (-i)^{M+ \beta}
 e^{-i (N-M) \phi} 
 \left[ J_{M+ \beta -1} (r') + i e^{-i \phi} J_{M+ \beta} (r') \right],
\end{equation}
whose solution is represented as an integral form:
\begin{eqnarray}
\psi_1 & = & A_1 
 \int_{0}^{r'} dr' e^{i r' \cos \phi}
 \left[ J_{M+ \beta -1} (r') + i e^{-i \phi} J_{M+ \beta} (r') \right]
\nonumber \\
& \equiv & A_1 \left[ I_{11} - I_{12} \right],
\end{eqnarray}
where
\begin{eqnarray}
A_1 & = & \frac{1}{2} (-i)^{M + \beta} e^{- i (N-M) \phi} e^{-i r' \cos \phi},
\\
I_{11} & = & \int_{0}^{\infty} dr' e^{i r' \cos \phi}
 \left[ J_{M+ \beta -1} (r') + i e^{-i \phi} J_{M+ \beta} (r') \right]
\end{eqnarray}
and
\begin{equation}
I_{12} =  \int_{r'}^{\infty} dr' e^{i r' \cos \phi}
 \left[ J_{M+ \beta -1} (r') + i e^{-i \phi} J_{M+ \beta} (r') \right].
\end{equation}
For $\psi_2$, we take
\begin{equation}
\psi_2 \equiv A_2 \left[ I_{21} - I_{22} \right].
\end{equation}
Here, $A_2$, $I_{21}$ and $I_{22}$ are obtained by replacing 
$N \rightarrow -N-1$, $\beta \rightarrow 1- \beta$ and $\phi \rightarrow 
- \phi$ in $A_1$, $I_{11}$ and $I_{12}$, respectively. The first integrals
in $\psi_1$ and $\psi_2$, i.e., $I_{11}$ and $I_{21}$ can be calculated 
exactly:
\begin{eqnarray}
A_1 I_{11} & = & e^{-i r' \cos \phi - i \alpha \phi} \theta (\phi),
\nonumber \\
A_2 I_{21} & = & e^{-i r' \cos \phi - i \alpha \phi} \theta (- \phi),
\label{incidentwave}
\end{eqnarray}
where $\theta (x)$ is the usual step function.
The sum of above terms gives exactly the same incident wave with
that obtained by AB\cite{aharonov59}. That is, this result is independent of
finite $M$. Therefore, we can infer that only the infinite angular momentum
quantum states yield the incident wave.

\indent For $I_{12}$ and $I_{22}$, we follow the similar procedure to AB
with finite $M$, in order to obtain the scattering wave. Then, in the
asymptotic limit, we obtain the followings:
\begin{eqnarray}
A_1 I_{12} & = & \frac{1}{2 \sqrt{2 \pi}} (-i)^{M + \beta} e^{-i (N-M) \phi}
\nonumber \\
& \times & \frac{1}{ \sqrt{r'}} \left\{ (-i)^{M + \beta - \frac{3}{2}}
 \frac{e^{ir'}}{1+ \cos \phi} (1+ e^{-i \phi}) + i^{ M+ \beta - \frac{3}{2}}
 \frac{e^{-ir'}}{1- \cos \phi} (1- e^{-i \phi}) \right\}
\label{psi1-2}
\end{eqnarray}
and
\begin{equation}
A_2 I_{22} = A_1 I_{12} ~[ N \rightarrow -N-1, ~\beta \rightarrow 1- \beta,~
 \phi \rightarrow - \phi ].
\label{psi2-2}
\end{equation}

\indent Now we calculate the last term $\psi_3$ in the asymtotic limit:
\begin{eqnarray}
\psi_3 & = & \frac{1}{\sqrt{2 \pi r'}} 
 \bigg\{ e^{ir'} \frac{1}{1+ e^{i \phi}} 
 \left( 1- (-1)^M e^{iM \phi} \right) e^{-i N \phi}
 \left[ (-i)^{\frac{1}{2} - 2 \beta - 2M} e^{-iM \phi} +
 (-i)^{\frac{1}{2} + 2 \beta} \right] 
\nonumber \\
 & + &  e^{-ir'} i^{\frac{1}{2}} \frac{e^{-i (N-M) \phi}}{1- e^{i \phi}}
 (1 - e^{2iM \phi}) \bigg\}.
\label{psi3}
\end{eqnarray}
Finally, adding Eqs.(\ref{incidentwave}), (\ref{psi1-2}), (\ref{psi2-2}) 
and (\ref{psi3}), we obtain the incident and scattered wave which is 
independent of $M$:
\begin{eqnarray}
\psi & = & \psi_1 + \psi_2 + \psi_3
\nonumber \\
& = & e^{-i r' \cos \phi - i \alpha \phi} - 
 \frac{i}{\sqrt{2 \pi i r'}} \frac{e^{ir'}}{\cos \frac{\theta}{2}}
 e^{-i \frac{\phi}{2}} e^{-i N \phi} \sin \beta \pi.
\label{finalresult}
\end{eqnarray}
As we have shown above, the incident wave is obtained from $A_1 I_{11} +
A_2 I_{21}$ and is independent of $\psi_3$. We may choose $M$ large enough
but finite, so that $\psi_3$ is most of the contributions in summation and
$\psi_1$ and $\psi_2$ contain only the infinite angular momenta. 
Since the incident wave are represented only by $\psi_1$ and
$\psi_2$, the information about the incident wave is involved in the
$m= \pm \infty$. In classical mechanics, the large angular momentum
in scattering corresponds to the large impact parameter from the scatterer
lacated at the origin when velocity is constant. Since AB potential is 
long-ranged, it is natural that only the particles infinitely far away from
the flux tube keep the information about the incident wave, and all the 
particles within a finite range are scattered. So we expect that this result 
may be a general property for any long-ranged potential scattering problems.
This also can be inferred from the fact that in the $\delta$-function 
potential, the infinitely short-ranged potential, only the $s$-wave has
an information on the scattering such as phase shift.\cite{jackiw91}

\indent 
\indent In conclusion, from the fact that the infinite angular momentum states
correspond exactly to the pole in the integral representation of wave function,
we find that the incident wave in AB scattering is obtained from the only
infinite angular momentum states. It is not certain, however, whether this 
property holds true for all the long-ranged interaction problems.
To ascertain whether this is a general property,  
some scattering problems other than AB scattering
must be investigated. This is now in progress.

\end{document}